\documentclass[10pt,conference]{IEEEtran}
\usepackage{cite}
\usepackage{amsmath,amssymb,amsfonts}
\usepackage{algorithmic}
\usepackage{graphicx}
\usepackage{textcomp}
\usepackage{xcolor}
\usepackage{comment}
\usepackage{enumitem}
\usepackage{tikz}
\usetikzlibrary{shapes, arrows.meta, positioning, calc}
\usepackage[margin=1in]{geometry}
\usepackage{booktabs}
\usepackage{hyperref}

\def\BibTeX{{\rm B\kern-.05em{\sc i\kern-.025em b}\kern-.08em
    T\kern-.1667em\lower.7ex\hbox{E}\kern-.125emX}}

\begin{document}

\title{From Prototype to Classroom: An Intelligent Tutoring System for Quantum Education}

\author{\IEEEauthorblockN{Iizalaarab Elhaimeur}
\IEEEauthorblockA{\textit{Center for Real-Time Computing} \\
\textit{Computer Science Department}\\
\textit{Old Dominion University}\\
Norfolk, VA \\
ielha003@odu.edu}
\and
\IEEEauthorblockN{Nikos Chrisochoides}
\IEEEauthorblockA{\textit{Center for Real-Time Computing} \\
\textit{Computer Science and Physics Departments}\\
\textit{Old Dominion University}\\
Norfolk, VA \\
nikos@cs.odu.edu}
}

\maketitle

\begin{abstract}
Quantum computing instructors face a compounding problem: the concepts are counterintuitive, the mathematical formalism is dense, and qualified faculty are scarce outside a small number of well-resourced institutions. Our prior work introduced a knowledge-graph-augmented tutoring prototype with two specialized LLM agents: a Teaching Agent for dynamic interaction and a Lesson Planning Agent for lesson generation. Validated on simulated runs rather than in a real course, that prototype left open whether more aggressive agent specialization would be needed to handle the full range of quantum education tasks under real student load. This paper answers the three questions that the prototype could not answer. Can agent specialization solve the reliability problem in a domain as technically demanding as quantum information science? Can the system run in a real course, not a demonstration? Does the instructor gain actionable intelligence from the deployment? We present ITAS (Intelligent Teaching Assistant System), a multi-agent tutoring system built around four contributions: a five-module QIS curriculum grounded in Watrous's information-first framework, a Spoke-and-Wheel teaching architecture with quantum-specialized  agents, a cloud infrastructure designed for production use and regulatory compliance, and a conversational analytics layer for instructors and content developers. Piloted in a quantum computing course at Old Dominion University, the system supports all three answers: deployment evidence is consistent with specialization addressing the task-boundary failures observed in the prototype, cloud infrastructure supports classroom-scale concurrency at sub-textbook cost, and the analytics agent surfaces curriculum gaps the instructor could not otherwise see.
\end{abstract}


\section{Introduction}
\label{sec:intro}

Quantum information science education has a faculty access problem. Qualified instructors are concentrated at a small number of research-intensive institutions, and the compensation gap between academia and industry makes hiring difficult everywhere else~\cite{fox2020revolution,meyer2024disparities}. As quantum computing transitions from research curiosity to industrial capability~\cite{ibmquantum,nqi2018}, the bottleneck is not curriculum or hardware access; it is the scarcity of people who can teach the subject at scale.

Our prior work addressed this directly, introducing a dual-agent tutoring prototype for graduate QIS education~\cite{elhaimeur2025toward}. That system demonstrated that LLM-based tutoring could engage with quantum formalism, explain Dirac notation, and respond to Qiskit errors. It also exposed a fundamental design failure: a single agent responsible for video comprehension, code debugging, and conceptual guidance simultaneously hallucinated at task boundaries, conflating timestamps with function names and mixing abstraction levels. The prototype answered whether AI tutoring for QIS was viable in principle. It left three questions open.

\textbf{Q1:} Does agent specialization solve the reliability problem in a domain as technically demanding as quantum information science?

\textbf{Q2:} Can the system run in a real course, not a controlled demonstration?

\textbf{Q3:} Does the instructor and content developer gain actionable intelligence from the deployment, without compromising student privacy?

This paper answers all three through the design and deployment of ITAS (Intelligent Teaching Assistant System). The system makes four contributions, each directly targeting one of the open questions or establishing the foundation that all three require.

\begin{enumerate}[leftmargin=*,noitemsep,topsep=0pt]
    \item \textbf{Five-Module QIS Curriculum.} A structured curriculum built on Watrous's information-first framework~\cite{watrous2025understandingquantuminformationcomputation}, with checkpoint exercises enforcing both correct output and correct implementation approach. The curriculum is the foundation on which the remaining contributions operate.

    \item \textbf{Spoke-and-Wheel Teaching Architecture (Q1).} Three parallel specialist agents, each with quantum-specific domain knowledge, coordinated by a Synthesizer. Specialization addresses the task-boundary hallucinations of the prototype: no agent reasons outside its designated scope.

    \item \textbf{Cloud Migration and Scalability (Q2).} Migration from a local prototype to Google Cloud Platform, delivering scalable, reliable, and secure infrastructure for real classroom deployment. Supporting analysis in~\cite{elhaimeur2026latency} confirms flat sub-4-second response times at classroom-scale concurrency and per-student cost below a STEM textbook.

    \item \textbf{Conversational Instructor Analytics Layer (Q3).} A single-agent analytics system that enables instructors to interrogate deployment data through natural language, surfacing aggregate pedagogical patterns without exposing individual student interactions.
\end{enumerate}

A note on scope: this paper focuses exclusively on the execution layer - covering real-time teaching, student interaction, and instructor feedback. The planning layer from the prior prototype, including the knowledge graph and adaptive curriculum sequencing, is deliberately deferred to future work. The generic multi-agent architecture is detailed in~\cite{elhaimeur2026architecture}; latency and scalability analysis appear in~\cite{elhaimeur2026latency}.

\section{Related Work}
\label{sec:related}

This section covers the four bodies of work that directly inform ITAS's design decisions and findings. Each citation is revisited in the section where it is applied.

\subsection{Intelligent Tutoring Systems}

AutoTutor established that conversational scaffolding, including hints, targeted feedback, and error correction, produces measurable learning gains through natural dialogue~\cite{graesser2004autotutor}. LLM-based tutoring has since extended this model across STEM disciplines~\cite{chu2025llmagents}, though evaluations have typically covered short sessions rather than sustained semester-long deployment with hands-on exercises.

\subsection{Multi-Agent LLM Architectures}

Kim et al.\ provide the quantitative foundation for ITAS's architectural choices~\cite{kim2025scaling}, reporting that centralized coordination improves performance by 80.9\% on parallelizable tasks while containing error amplification to 4.4$\times$ compared to 17.2$\times$ for independent agents. AutoGen demonstrates role-based conversable agents producing structured JSON outputs~\cite{wu2024autogen}, and MetaGPT shows that standardized operating procedures per agent reduce cascading hallucination errors~\cite{hong2024metagpt}. These three works are the direct precedents for ITAS's Spoke-and-Wheel design.

\subsection{Quantum Computing Education}

The workforce demand for QIS-literate graduates continues to exceed supply~\cite{fox2020revolution}. ITAS adopts Watrous's information-theoretic framework, which builds quantum concepts as generalizations of classical information processing~\cite{watrous2025understandingquantuminformationcomputation,JW-IBM_Videl-Lectures}. Student misconceptions in this approach are documented by Hu et al.~\cite{hu2024qcmisconceptions} and Majidy~\cite{majidy2025misconceptions}, and their findings directly inform the sections on LLM capability limits. Galetto et al.\ validate embedding executable code exercises alongside lecture content~\cite{galetto2024labs}, the model ITAS extends with real-time AI tutoring.

\subsection{Learning Analytics}

Guo et al.\ establish video seek behavior and drop-off timing as reliable proxies for content engagement~\cite{guo2014video}. Blikstein et al.\ demonstrate that high-frequency code telemetry reveals behavioral archetypes and that combining heterogeneous data streams enables cross-modal insights no single stream can provide~\cite{blikstein2014programming,blikstein2016multimodal}. Kizilcec et al.\ show that clustering interaction patterns surfaces prototypical learner subpopulations~\cite{kizilcec2013archetypes}. These three methodologies underpin the Deployment and Analysis section.

\section{QIS Curriculum Design}
\label{sec:curriculum}

The curriculum is the foundation on which the teaching architecture, infrastructure, and analytics layer all operate. Two design choices make it a contribution in its own right. First, building on Watrous's information-first sequencing, each module builds on the previous through explicit mathematical generalization, enabling the teaching agents to support cross-module synthesis rather than treating each lesson as isolated content. A topic-first curriculum organized around gates and algorithms rather than information-theoretic concepts does not have this property. Second, the checkpoint philosophy enforces implementation approach alongside output correctness, which matters specifically in quantum education where a student can reach a correct measurement result through an implementation that reflects no understanding of the underlying formalism.

\subsection{Framework and Module Structure}

The curriculum follows Watrous's information-first approach to quantum computing~\cite{watrous2025understandingquantuminformationcomputation,JW-IBM_Videl-Lectures}, which treats quantum concepts as generalizations of classical information processing rather than departures from it. The five modules are structured around two pedagogical pillars: learning-by-doing through exercises that reinforce the material and project-driven learning, where students are able to connect QIS to their own interests.

\textbf{Module 1: Single Systems.} Establishes the classical information foundation before transitioning to quantum. Students work with Dirac notation, represent quantum states as complex vectors, and apply unitary operations including Pauli gates and the Hadamard gate.

\textbf{Module 2: Multiple Systems.} Extends to compound systems through tensor products, covering independence, correlation, and entanglement as properties of compound states, with operations on multi-qubit systems including CNOT and controlled gates.

\textbf{Module 3: Quantum Circuits.} Introduces the circuit model. Students construct circuits for Bell states, GHZ states, and the quantum Fourier transform, and encounter fundamental limitations including the no-cloning theorem and state discrimination.

\textbf{Module 4: Entanglement in Action.} Applies entanglement as a computational resource through quantum teleportation, superdense coding, and the CHSH game, connecting to Bell's theorem and the 2022 Nobel Prize in Physics.

\textbf{Module 5: Quantum Circuit Cutting.} Introduces circuit cutting 
and knitting as near-term techniques for executing circuits on hardware with 
limited qubit counts~\cite{oliver2025capstone,tang2025enabling}, bridging 
the gap between the theoretical circuit model and practical hardware 
constraints. This module anchors the project-driven learning pillar, where 
students apply circuit cutting to problems in their own disciplines, as 
demonstrated by recent work in hybrid quantum-classical optimization for 
vehicle routing~\cite{maciejunes2025vrp} and quantum edge detection for 
medical imaging~\cite{billias2025qhed}.

\subsection{Checkpoint Design Philosophy}

Each checkpoint enforces both correct output and correct implementation approach. This design choice is specific to quantum education, where multiple valid implementations may produce identical measurement statistics while reflecting fundamentally different levels of understanding. Unlike platforms that enforce a single predetermined solution path, ITAS checkpoints accept any valid implementation satisfying both criteria simultaneously. Across the deployment, students reached correct answers via genuinely different implementation paths, confirming that the checkpoint design preserves exploratory learning while maintaining pedagogical rigor.

Each checkpoint carries three separate agent instruction sets. Autograding instructions specify exact pass and fail criteria with the directive that correct output alone is not sufficient. Guidance instructions frame the correct approach as a series of hints, with an explicit prohibition on writing complete solutions. Debugging instructions include a catalog of common errors per checkpoint to enable targeted diagnosis. This triple-instruction design provides differentiated scaffolding at the verification, conceptual, and error-diagnosis levels without solving problems for students~\cite{graesser2004autotutor,galetto2024labs,wootton2021qiskit}.

\section{Spoke-and-Wheel Teaching Architecture}
\label{sec:architecture}

The teaching architecture is the direct response to the prototype's core failure. This section describes the design rationale, the quantum-specific specialization decisions, and the analytics architecture that complements the teaching system. The generic multi-agent design is detailed in~\cite{elhaimeur2026architecture}.

A note on scope: the prior prototype included a knowledge graph designed for curriculum sequencing and adaptive lesson planning. That planning layer is deliberately out of scope here. The execution layer, covering real-time teaching and immediate feedback, is the focus of this paper, and session-based state in a cloud-hosted relational database proved sufficient for it. 

\subsection{From Prototype to Specialization}

The prior dual-agent system left the Teaching Agent overloaded, responsible for video comprehension, code debugging, and conceptual guidance simultaneously. Hallucination appeared at task boundaries: the agent suggested nonexistent timestamps when reasoning about code, and mixed abstraction levels when moving between mathematical formalism and implementation~\cite{huang2024hallucination}. The root cause was cognitive overload, not model capability.

The Spoke-and-Wheel architecture addresses this through task-based decomposition. Three specialist agents analyze each student question in parallel, each reasoning only within its designated scope. A Synthesizer Agent integrates their outputs using a priority hierarchy, with code errors addressed first, then conceptual gaps, then video references. The three-agent count emerged empirically; two agents conflated code and concepts, while four added coordination overhead without reliability improvement. Across 75 student interactions spanning seven question categories, the authors' review of the logged exchanges identified no task-boundary failures of the type documented in the prototype. A controlled, blinded comparison against the prototype on the same input set remains future work. These observations are consistent with Kim et al.'s finding that centralized coordination substantially improves performance on parallelizable tasks while containing error amplification relative to independent agents~\cite{kim2025scaling}.

Specialization reduces hallucination in three ways. Cognitive load reduction ensures each agent reasons about one task only. Task interference elimination means the Video Agent cannot hallucinate function names because it never reasons about code. Synthesis filtering allows the Synthesizer to discard low-confidence outputs before they reach the student.

\begin{figure}[h]
\centering
\includegraphics[width=\columnwidth]{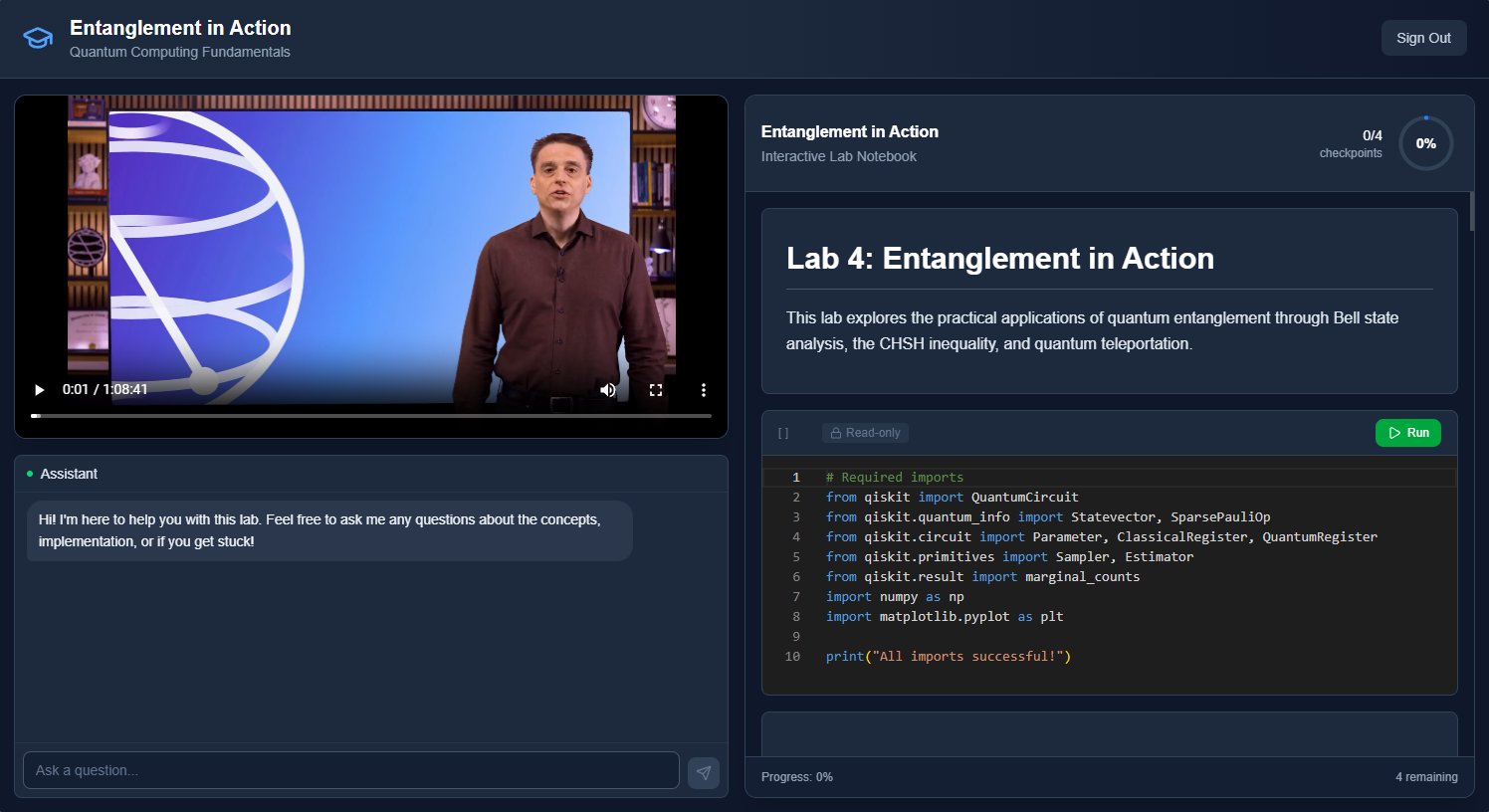}
\caption{The ITAS student interface. Three panels (video player, interactive Qiskit notebook, and AI teaching agent chat) are not merely co-located but actively interact. A code error in the notebook triggers the Code Agent's diagnosis, which the Synthesizer cross-references with the Video Agent's timestamp map to surface the relevant lecture segment in the video panel.}
\label{fig:interface}
\end{figure}

\subsection{Quantum-Specific Agent Design}

Each agent carries domain knowledge specific to QIS instruction. General-purpose prompting is insufficient; the specialization must be grounded in the content the students encounter.

The \textbf{Video Agent} receives a timestamp-indexed concept map for each Watrous lecture, enabling targeted video references rather than generic suggestions to rewatch. The \textbf{Guidance Agent} is calibrated with an explicit directive: ``These are graduate students. Engage with the mathematics; do not oversimplify.'' The \textbf{Code Agent} carries a Qiskit error catalog compiled from the most common failure modes in the course material, covering operator misuse, deprecated API imports, normalization errors, and the big-endian qubit ordering conventions specific to Qiskit. The catalog was hand-curated by the authors before deployment, drawing on common Qiskit error patterns from prior course offerings and the official Qiskit textbook~\cite{wootton2021qiskit}; it was not derived from the deployment data itself. The ability to redirect students from deprecated APIs to current ones proved particularly valuable as Qiskit versions evolved during the deployment.

The triple-instruction design described in Section~\ref{sec:curriculum} is an architectural decision as much as a pedagogical one. It partitions the agent's responsibilities at the instruction level, ensuring that autograding, guidance, and debugging never interfere with one another within a single agent call.

\begin{figure}[h]
\centering
\begin{tikzpicture}[
    scale=0.62, transform shape,
    node distance=1.1cm and 1.4cm,
    every node/.style={font=\small},
    block/.style={rectangle, draw, rounded corners=3pt,
                  minimum height=0.75cm, minimum width=2.0cm,
                  align=center, fill=#1!12, draw=#1!60, line width=0.6pt},
    block/.default=blue,
    arrow/.style={-{Stealth[length=5pt]}, thick, color=#1!70},
    arrow/.default=black,
]
\node[block=gray, minimum width=2.8cm] (query) {Student Query};
\node[block=red,  below left=1.1cm and 1.6cm of query] (video)    {Video Agent};
\node[block=blue, below=1.1cm of query]                 (guidance) {Guidance Agent};
\node[block=green,below right=1.1cm and 1.6cm of query] (code)     {Code Agent};
\node[block=purple,below=1.1cm of guidance, minimum width=2.8cm] (synth) {Synthesizer Agent};
\node[block=gray, below=1.0cm of synth,    minimum width=2.8cm] (response) {Student Response};
\draw[arrow=gray] (query.south) -- ++(0,-0.25) -| (video.north);
\draw[arrow=gray] (query.south) -- (guidance.north);
\draw[arrow=gray] (query.south) -- ++(0,-0.25) -| (code.north);
\draw[arrow=red]  (video.south)    |- (synth.west);
\draw[arrow=blue] (guidance.south) -- (synth.north);
\draw[arrow=green](code.south)     |- (synth.east);
\draw[arrow=purple](synth.south)   -- (response.north);
\node[font=\scriptsize\itshape, color=gray!80,
      left=0.55cm of video] {parallel};
\end{tikzpicture}
\caption{The Spoke-and-Wheel teaching architecture. Three specialist agents execute in parallel; the Synthesizer waits for all three before producing a response. End-to-end latency is dominated by $\max(L_{\text{video}},\,L_{\text{guidance}},\,L_{\text{code}}) + L_{\text{synth}}$.}
\label{fig:architecture}
\end{figure}
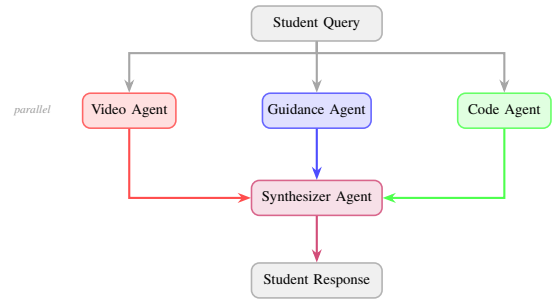

Figure~\ref{fig:architecture} shows the Spoke-and-Wheel design. A representative interaction illustrates the coordination: a student encounters \texttt{QiskitError: 'other is not a number'} when applying Pauli-Z to $|1\rangle$ and asks what the error means. The Video Agent identifies a relevant lecture segment but determines the error is implementation-specific. The Code Agent identifies the use of Python's \texttt{*} operator instead of \texttt{numpy.dot}, which Qiskit's \texttt{Operator} class does not support for element-wise multiplication with \texttt{Statevector}. The Guidance Agent confirms the student's conceptual model is intact. The Synthesizer validates understanding, pinpoints the operator issue, and suggests correct syntax without writing complete code. The student receives a response that addresses video context, code diagnosis, and conceptual confirmation simultaneously.

\subsection{Analytics Architecture}

The analytics system uses a deliberately simpler design than the teaching architecture. A single conversational agent queries pre-aggregated BigQuery summaries~\cite{blikstein2016multimodal}, assembles results with lesson metadata, and produces a natural language response. The agent narrates pre-queried data and never writes SQL or performs calculations. This narrow-scope design applies the same reliability principle as the teaching architecture: an agent that cannot exceed its designated function cannot hallucinate beyond it.

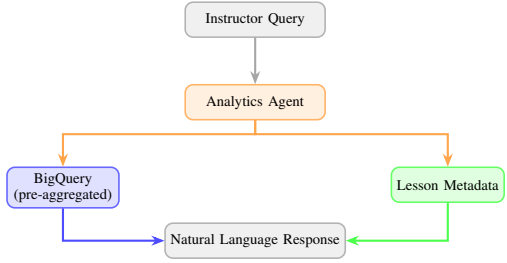
\begin{figure}[h]
\centering
\begin{tikzpicture}[
    scale=0.62, transform shape,
    node distance=1.0cm and 1.2cm,
    every node/.style={font=\small},
    block/.style={rectangle, draw, rounded corners=3pt,
                  minimum height=0.75cm, minimum width=2.4cm,
                  align=center, fill=#1!12, draw=#1!60, line width=0.6pt},
    block/.default=blue,
    arrow/.style={-{Stealth[length=5pt]}, thick, color=#1!70},
    arrow/.default=black,
]
\node[block=gray,   minimum width=3.0cm] (query)    {Instructor Query};
\node[block=orange, below=1.0cm of query, minimum width=3.0cm] (agent)
      {Analytics Agent};
\node[block=blue,   below left=1.0cm and 1.4cm of agent]  (bq)
      {BigQuery\\(pre-aggregated)};
\node[block=green,  below right=1.0cm and 1.4cm of agent] (meta)
      {Lesson Metadata};
\node[block=gray,   below=2.2cm of agent, minimum width=3.0cm] (response)
      {Natural Language Response};
\draw[arrow=gray]   (query.south)  -- (agent.north);
\draw[arrow=orange] (agent.south)  -- ++(0,-0.3) -| (bq.north);
\draw[arrow=orange] (agent.south)  -- ++(0,-0.3) -| (meta.north);
\draw[arrow=blue]   (bq.south)    |- (response.west);
\draw[arrow=green]  (meta.south)  |- (response.east);
\end{tikzpicture}
\caption{The analytics architecture. A single conversational agent queries pre-aggregated BigQuery data and lesson metadata, then produces a natural language response for the instructor. Narrow scope is the reliability principle applied to analytics.}
\label{fig:analytics_arch}
\end{figure}

Figure~\ref{fig:analytics_arch} shows the analytics pipeline. The design contrast with the Spoke-and-Wheel system is intentional: the teaching problem required parallel specialization; the analytics problem required a single agent with constrained scope.

\section{Cloud Migration and Scalability}
\label{sec:infrastructure}

The prototype in \cite{elhaimeur2025toward} ran locally. Local execution cannot support a real classroom: there is no concurrent request handling, no fault tolerance, no session persistence across network boundaries, and no security boundary between student code execution and the host environment. The migration to Google Cloud  Platform (GCP), supported by Monarch Sphere~\cite{monarchsphere}, is the direct solution to these constraints, and it is a contribution in its own right. This section describes the migration, its outcomes, and the evidence that the resulting system supports classroom-scale deployment.

\subsection{Migration Outcomes}

The migration delivers three outcomes, each addressed by specific infrastructure decisions.

\begin{figure}[h]
\centering
\includegraphics[width=\columnwidth]{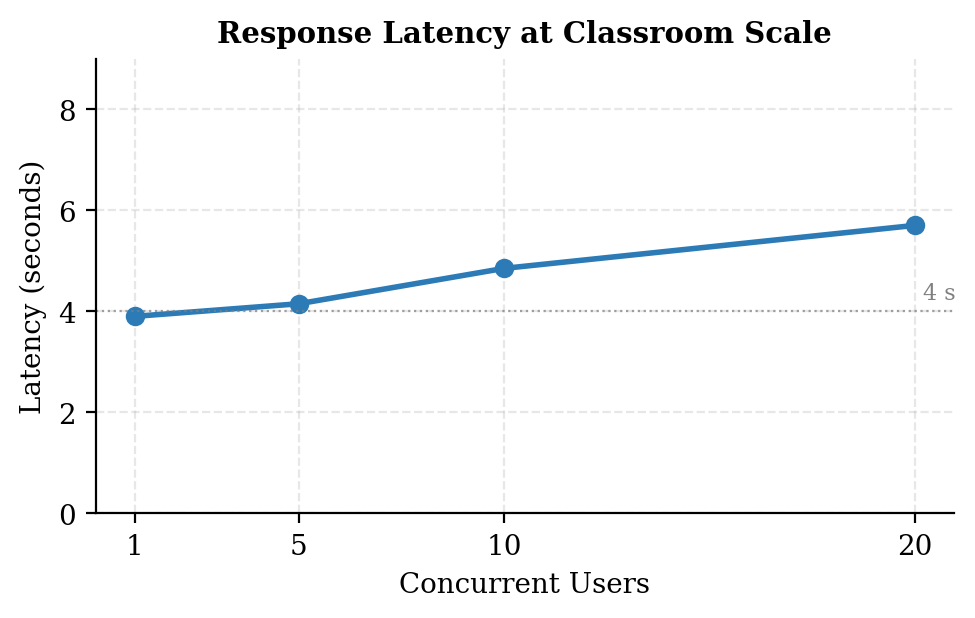}
\caption{Estimated median end-to-end response latency vs.\ concurrent users at classroom scale (up to 20 simultaneous users). Latency remains manageable across the seminar-to-classroom range. Full analysis across deployment scales and infrastructure configurations appears in~\cite{elhaimeur2026latency}.}
\label{fig:latency}
\end{figure}

\textbf{Scalability.} Four Cloud Run microservices handle concurrent classroom load with auto-scaling and minimum instance configuration to eliminate cold start degradation. Analysis in~\cite{elhaimeur2026latency} confirms flat response latency at classroom-scale concurrency, covering the seminar-to-classroom range without manual capacity planning. Figure~\ref{fig:latency} shows the estimated latency curve across this range.

\textbf{Reliability.} Analytics events stream via Pub/Sub to BigQuery in fire-and-forget mode~\cite{hohpe2003eip}. This decouples the teaching pipeline from data collection: a failure in the analytics ingestion path does not affect teaching agent responses, and a teaching agent failure does not corrupt the analytics record.

\textbf{Privacy, Security and Compliance} Student code executes in a container-level sandboxed environment, preventing any student submission from affecting the host system or other students' sessions. The deployment complies with the Family Educational Rights and Privacy Act (FERPA)~\cite{ferpa1974}, which prohibits disclosure of personally identifiable information from student education records without consent. Compliance is enforced through three mechanisms: all interaction data is stored under anonymized identifiers, the analytics agent exposes only aggregate pedagogical patterns rather than individual student records, and all data remains within Old Dominion University's controlled GCP environment under institutional data governance policies.

\subsection{Microservice Design}

The system runs as four Cloud Run microservices, each responsible for one function. The Teaching Agent Service hosts the Spoke-and-Wheel agents with auto-scaling. The Python Execution Service provides sandboxed code execution pre-loaded with quantum computing dependencies. The Analytics Ingestion Service streams events to BigQuery. The Autograding Service evaluates checkpoint submissions, verifying both the correctness of the output and the validity of the implementation approach.

One deployment finding is worth noting explicitly. A missing \texttt{pylatexenc} dependency for circuit visualization caused a cluster of errors in Module 3 before resolution. This kind of failure is only visible in real deployment, not in a controlled demonstration, and it illustrates the gap between a working prototype and production infrastructure.

\subsection{Cost at Classroom Scale}

\begin{figure}[h]
\centering
\includegraphics[width=\columnwidth]{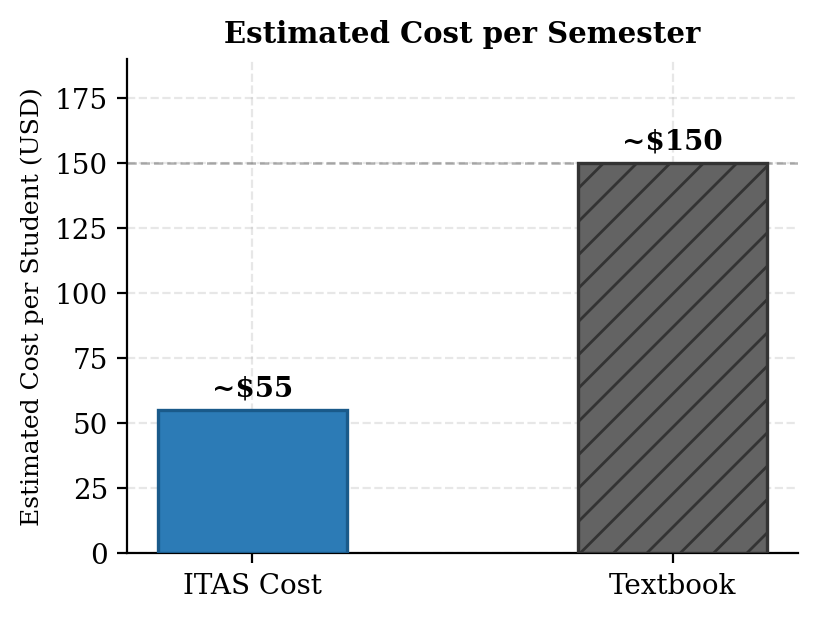}
\caption{Estimated per-student cost per semester compared to a typical STEM textbook. The estimate reflects a worst-case usage ceiling of 10,000 queries per student per semester. Actual costs will be lower under realistic usage. Full cost modeling and infrastructure configuration details appear in~\cite{elhaimeur2026latency}.}
\label{fig:cost}
\end{figure}

Figure~\ref{fig:cost} shows estimated per-student cost at classroom scale compared to a typical STEM textbook. Under realistic infrastructure configurations, cost remains well below the textbook baseline even at a conservative usage estimate, with pay-per-use pricing ensuring that cost scales with actual usage rather than peak capacity reservation. Full cost and concurrency analysis across deployment scales appears in~\cite{elhaimeur2026latency}.

\section{Conversational Instructor Analytics}
\label{sec:analytics}

When students interact with ITAS, the system accumulates detailed knowledge of each student's trajectory. The instructor sees almost none of this by design. The student-AI interaction space is intentionally protected as a low-stakes environment for exploration and error, a space where students can ask questions they might not raise in class, without those questions appearing in a gradebook or instructor dashboard. We call the resulting situation the Blind Instructor Problem: the system that most needs to inform the instructor is the one least able to expose its data without compromising the conditions that made the data valuable.

The conversational analytics agent addresses this by providing aggregate pedagogical intelligence without individual exposure. The instructor queries patterns; the system never returns individual transcripts or identifiable interaction logs. All findings reported through the agent represent synthesized observations across the full deployment, not traceable to any single student.

\subsection{Capabilities}

The analytics agent operates through a three-stage pipeline described in Section~\ref{sec:architecture}. Its value lies not in the pipeline but in what the pipeline enables: cross-referencing three data streams simultaneously in response to a natural language question. A static dashboard can show video drop-off rates. It cannot explain why students stopped watching at a specific timestamp, correlate that drop-off with the absence of exercises covering that content, and present the finding to the instructor as a curriculum revision recommendation.

A representative session from the pilot deployment illustrates the capability gap - what we term the \textbf{dead zone finding}. Asked about student engagement in Module 2, the analytics agent identified a cluster of students who stopped watching the lecture around the 42-minute mark. Cross-referencing the timestamp with the exercise coverage map, the agent identified that all four checkpoints in Module 2 cover single-qubit operations, while the lecture content from 44 minutes onward covers multi-qubit states, entanglement, and all four Bell states. Students watched up to the point where they had enough background to complete the exercises, then disengaged. The agent surfaced this finding without an explicit instructor query about it. The instructor used the finding to plan targeted in-class follow-up and to identify the curriculum revision needed to close the gap.

The agent also demonstrated the ability to distinguish types of failure. A checkpoint submission using an incorrect variable name was identified as a reading comprehension issue, distinct from a \texttt{QiskitError} in another submission that reflected a genuine conceptual gap. This distinction matters for instructor intervention: the first calls for a clarification note, the second calls for a lecture revision.

\begin{figure}[h]
\centering
\includegraphics[width=\columnwidth]{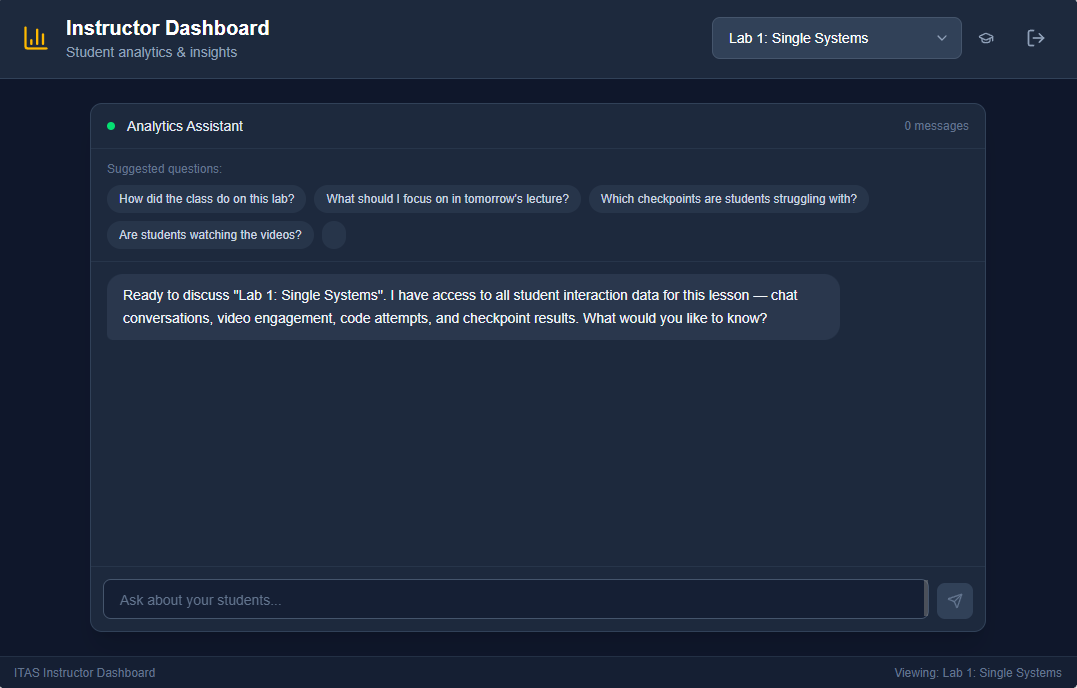}
\caption{The ITAS instructor analytics interface. The conversational agent accepts natural language queries and cross-references video seek data, code execution logs, and checkpoint submissions stored in BigQuery to surface aggregate pedagogical patterns. The interface exposes patterns, not individuals.}
\label{fig:analytics}
\end{figure}

\section{Deployment and Analysis}
\label{sec:findings}

ITAS was deployed in a pilot classroom setting at Old Dominion University as part of a quantum computing course. In keeping with the privacy design described in Section~\ref{sec:analytics}, this section reports only aggregate observations and behavioral archetypes. No individual student details or identifiable interaction patterns are disclosed.

\subsection{Evidence Base}

The deployment involved 5 graduate students enrolled in the course over a single semester. It generated interaction events across five categories: video playback, code execution, chat messages, checkpoint submissions, and session management. The distribution across modules reflects a pattern of progressive attrition across the core curriculum, with fewer students reaching each successive module. A mid-semester re-engagement was observed when Module 5 was introduced, suggesting that new content can recover students who had paused the self-paced track.

\subsection{What the Data Showed}

The following three subsections address each open question in turn, drawing on chat transcripts, code execution logs, video engagement patterns, and checkpoint submissions as evidence. Each subsection states the answer before presenting the evidence that supports it.

\subsubsection{Q1: Does AI Agent Specialization Solve the Reliability Problem?}

The deployment evidence is consistent with yes, though a controlled comparison against the prototype remains future work. The domain demands simultaneous competence across three distinct tasks that no single agent can hold without sacrificing depth in at least one, and the specialized architecture handled all three without the task-boundary failures observed in the prototype \cite{elhaimeur2025toward}. Table~\ref{tab:categories} shows the distribution of question categories across the chat data. The dominant categories are study strategies (23\%), meaning questions about how to approach the material, pacing, and learning resources rather than the content itself, conceptual quantum (20\%), and mathematical formalism (17\%), with implementation and Qiskit questions at 12\%. This distribution confirms why a single general-purpose agent fails: the domain requires simultaneous competence in mathematical formalism, implementation specifics, and video-indexed content, all at graduate level. No single agent prompt can hold all three without sacrificing depth in at least one.

\begin{table}[h]
\centering
\caption{Student Question Categories}
\label{tab:categories}
\begin{tabular}{lrc}
\toprule
\textbf{Category} & \textbf{Count} & \textbf{\%} \\
\midrule
Study Strategies         & 17 & 23\% \\
Conceptual Quantum       & 15 & 20\% \\
Mathematical Formalism   & 13 & 17\% \\
Implementation/Qiskit    & 9  & 12\% \\
Social/Exploratory       & 10 & 13\% \\
Video Reference          & 6  & 8\%  \\
Cross-Lesson Synthesis   & 5  & 7\%  \\
\bottomrule
\end{tabular}
\end{table}

The following exchange, condensed from a Module 1 session, illustrates the depth at which the system operated. The student had been working through deterministic operations and began probing the mathematical structure:

\smallskip
\noindent\textbf{Student:} ``Is the matrix unique because of certain properties that the function $f$ possesses or is it true in general?''

\noindent The Guidance Agent explained that for a function $f: S \to S$ on a finite set, the matrix representation is unique given a fixed basis. The student then generalized:

\noindent\textbf{Student:} ``What about if I consider instead of $\mathbb{R}$ a set $S$ with $|S| = n$, $n$ finite? What are then the properties of $f$?''

\noindent The agent described that the corresponding matrix is a permutation matrix if and only if $f$ is a bijection. The student then connected this to the lecture:

\noindent\textbf{Student:} ``John referred to associative operation in the context of Deterministic operations (30:48). Can you explain how this property is used?''
\smallskip

\noindent This sequence is characteristic of graduate QIS engagement: the student moves from specific to general, connects formal mathematics to a video timestamp, and drives toward deeper abstraction. The Video Agent located the relevant lecture segment; the Guidance Agent explained associativity in the context of function composition and matrix multiplication. This type of multi-turn, cross-task sequence represents exactly the failure mode documented in the prototype~\cite{elhaimeur2025toward}, where a single overloaded agent conflated video timestamps with function names and mixed abstraction levels when moving between mathematical formalism and implementation. Whether a single agent would have dropped the thread here cannot be claimed with certainty without a controlled comparison, which remains future work.

LLM-based tutoring handles several QIS tasks reliably. The agents provided accurate explanations of Dirac notation and its relationship to linear algebra, correct historical context for foundational concepts, and structured study materials synthesized from multi-turn conversations. Qiskit API guidance was correct in most cases, with the Code Agent successfully redirecting students from deprecated imports to current API patterns.

From our deployment experience, the agents performed less reliably on multi-qubit state verification, which requires executing code rather than reasoning about it. Circuit optimization similarly exposed agent limitations, as it demands awareness of gate costs and hardware topology that the agents do not carry. These observations are consistent with documented limitations of LLMs in quantum computing contexts~\cite{kharsa2024qc,hu2024qcmisconceptions}, where reasoning about hardware-specific constraints and executable verification have been identified as persistent failure modes. Version-specific API failures required infrastructure-level resolution rather than tutoring-level guidance, illustrating the boundary between what an LLM tutor can address and what requires a system administrator.

Checkpoint submission patterns confirm that the dual output-and-approach requirement functions as intended. Submissions were rejected not only for incorrect results but for correct results reached through implementations that did not satisfy the approach criteria, a distinction that would be invisible to a system checking outputs alone. In ITAS, this is enforced through the Autograding Agent, which receives per-checkpoint instructions specifying both the expected output and the criteria for acceptable implementations, rejecting hardcoded outputs or implementations that bypass the underlying concepts while accepting any solution that demonstrates genuine understanding. The agent evaluates submissions against both criteria simultaneously, returning a structured JSON response with a boolean pass flag and reasoning that separates the two criteria.

\subsubsection{Q2: Can This Run in a Real Classroom?}

Yes, at the scale of this deployment. The system ran, collected data across multiple modules over a semester, and produced behavioral patterns that only emerge from sustained deployment. A controlled demonstration cannot produce a late engager. A single-session test cannot produce a passive consumer detectable only through longitudinal data. The following archetypes are evidence of deployment, not just observation~\cite{kizilcec2013archetypes,blikstein2014programming}.

The \textbf{self-directed} archetype relied primarily on video and independent coding with minimal AI interaction, using the tutor primarily for error resolution. Characteristic errors reflected implementation knowledge gaps rather than conceptual misunderstanding, and the high code execution volume with low chat volume confirms that the system supports autonomous learners who prefer independent struggle.

The \textbf{tutor-reliant} archetype used the agent extensively for conceptual exploration, mathematical formalism, and cross-module synthesis, treating it as a collaborative study partner rather than a help-of-last-resort. This student pushed the agent toward formal definitions and group-theoretic connections in ways that extended beyond the conversational scaffolding model established by AutoTutor~\cite{graesser2004autotutor}, suggesting that graduate QIS learners may engage AI tutors as intellectual partners rather than remediation tools, a mode of interaction that existing Intelligent Tutoring System (ITS) frameworks do not fully account for.

The \textbf{late engager} archetype generated casual initial interactions before becoming a high-volume user in later modules, including a non-linear trajectory that skipped one module entirely and engaged deeply with subsequent ones. This pattern would be impossible to detect in a system enforcing sequential completion.

The \textbf{passive consumer} archetype engaged almost exclusively through video playback with no code execution and minimal chat interaction. This is a failure mode the reactive teaching system cannot address: students in this pattern never signal confusion. Only the analytics agent detected the disengagement.

The diversity of these trajectories, including non-linear module completion, tutor use as intellectual partner rather than help resource, and passive consumption invisible to the teaching system, is only visible because the deployment ran long enough and captured enough data to surface them.

Video engagement patterns confirmed the dead zone finding described in Section~\ref{sec:analytics}~\cite{guo2014video}. Code execution across the deployment produced 387 total executions with a 77\% overall success rate. Error analysis revealed three distinct categories: Qiskit API errors from operator misuse and deprecated imports, missing library errors concentrated in Module 3 before the \texttt{pylatexenc} dependency was resolved, and mathematical errors including tensor product shape mismatches and undefined variables. The Module 3 code execution success rate, defined as the percentage of executions that completed without throwing a runtime error, dropped to approximately 60\% before the infrastructure fix, recovering to match the 86--87\% rates observed in Modules 1 and 2 afterward. This connection between a deployment finding and an infrastructure decision is visible only because both data streams were captured~\cite{blikstein2016multimodal}.

\subsubsection{Q3: Does the Instructor Get Actionable Intelligence?}

Yes. The analytics agent surfaced two findings the instructor could not have obtained from any other source, without exposing any individual student data. The dead zone finding, described in detail in Section~\ref{sec:analytics}, is the clearest example: a structural curriculum gap identified automatically, without an explicit query, acted on mid-semester. What that section does not capture is the causal chain it set in motion. The instructor revised the Module 2 exercise set to include multi-qubit checkpoints, directly addressing the content coverage gap the agent identified. That revision was motivated entirely by the analytics agent's output, not by any student complaint or assessment result.

The passive consumer archetype provides a second answer. A student consuming over a thousand video events without writing code or asking a question would be far more likely to surface in a traditional in-person session, where physical presence and instructor observation provide at least some signal of disengagement, though even there such patterns can go undetected. In an AI-enhanced flipped classroom, that student is invisible without the analytics layer. The analytics agent made the pattern visible without exposing which student it was.

\subsection{What It Means}

Four design principles emerge from the deployment that apply to any instructor building or deploying an AI-assisted quantum computing course.

Exercise coverage must match content coverage. The dead zone finding demonstrates that students treat exercises as the operational definition of required material. Any quantum concept without a corresponding implementation exercise risks being skipped regardless of its importance~\cite{galetto2024labs,meyer2024content}.

Checkpoints should enforce implementation approach alongside correctness. In quantum computing, multiple implementations can produce identical measurement statistics. Requiring specific construction patterns ensures students develop the intended conceptual understanding and prevents hardcoding of correct outputs.

AI tutors for graduate QIS must match the audience's mathematical sophistication. Graduate students did not ask for simplified intuitive explanations; they asked for group-theoretic definitions and formal characterizations. Agent prompts calibrated to undergraduate audiences will underserve this population~\cite{meyer2022interdisciplinary}.

AI-enhanced flipped classrooms require more robust monitoring, not less. Students who would signal confusion in an in-person session may consume content passively in a self-paced environment. The analytics layer is not a convenience feature; it is a pedagogical necessity.

\subsection{Limitations}

The absence of a control group means we cannot attribute outcomes causally to ITAS rather than to the curriculum or instructor. The instructor designed both the curriculum and the system, introducing potential bias in both design and interpretation. Qualitative validity follows Patton~\cite{patton2002qualitative} and Merriam~\cite{merriam2009qualitative}: validity in qualitative inquiry depends on the information richness of cases and the triangulation of evidence sources, not on sample size. Controlled evaluation with larger cohorts and validated instruments~\cite{passante2023aceqis,wilcox2015upperdivision} remains important future work.

\section{Future Work}
\label{sec:future}

\textbf{Adaptive exercise generation.} Video seek patterns and chat confusion signals will drive automatic generation of exercises anchored to the timestamps where students disengage, directly addressing the dead zone finding.

\textbf{Proactive intervention.} Real-time engagement monitoring with instructor alerts triggered by passive consumption patterns, targeting the passive consumer archetype before disengagement becomes permanent.

\textbf{Bi-directional interactive lectures.} Rather than passive video consumption, this format replaces pre-recorded lectures with a live audio agent that delivers content through slides while allowing students to interrupt, ask questions, and resume seamlessly. Initial implementation covering Modules 4 and 5 has been tested in two formats: a one-on-one format where students interact with the agent at their own pace, and a whole-class format where the instructor and students prompt the agent together during class, with the agent functioning as a co-teacher that fields questions in real time. Both formats will be reported in the journal version of this work.

\textbf{Multi-language Code Agent.} Extending beyond Qiskit to PennyLane, Q\#, CUDA-Q, and diagrammatic languages such as Lambeq, each requiring its own error catalog and API conventions.

\textbf{Planning layer and knowledge graph.} Curriculum authoring, adaptive sequencing, and higher-order pedagogical strategy, the context where the prototype's knowledge graph will be revisited.

\textbf{Controlled evaluation.} AI-tutored versus human-TA comparison with pre-and-post assessments using validated instruments~\cite{passante2023aceqis,wilcox2015upperdivision} in larger course offerings.

\section{Conclusion}
\label{sec:conclusion}

This paper presented ITAS as a direct continuation of prior prototype work \cite{elhaimeur2025toward}, organized around three open questions that the prototype could not answer. The five-module QIS curriculum, grounded in Watrous's information-first framework, provided the pedagogical setting for all three answers. The Spoke-and-Wheel teaching architecture answered Q1: specialization by task addresses the hallucination at task boundaries that the prototype exhibited, confirmed by the deployment's checkpoint and chat data, and consistent with Kim et al.'s scaling principles~\cite{kim2025scaling}. The cloud migration answered Q2: production infrastructure on Google Cloud Platform supports classroom-scale concurrency at sub-textbook cost, with the gap between local prototype and real classroom bridged by scalable, reliable, and secure microservices. The conversational instructor analytics layer answered Q3: the Blind Instructor Problem is addressed through aggregate pattern surfacing that preserves student privacy while giving instructors the intelligence they need to act, confirmed by the dead zone finding and the passive consumer detection.

Deployment in a quantum computing course at Old Dominion University confirmed that all three answers hold outside a controlled demonstration. This is a pilot deployment in a precise sense: three things are absent from the current version that a production system would require. The evaluation is qualitative and without a control group, so learning outcomes cannot be attributed causally to ITAS. The deployment covered one course at one institution, so generalizability is not yet established. And the planning layer, including adaptive curriculum sequencing and the knowledge graph, is deferred. The current version demonstrates that the execution layer works. It does not yet demonstrate that it teaches better than the alternative. That question requires controlled evaluation, and it is the priority for future work.

\section*{Acknowledgments}
This research was sponsored in part by the Richard T. Cheng Endowment and supported by Monarch Sphere~\cite{monarchsphere}. Cloud infrastructure was provided through Google Cloud Platform research credits and John D. Pratt, Seth J. Hohensee, and Alex L. Tucker of the ITS Group at Old Dominion University. The QIS curriculum is based on John Watrous's IBM Quantum lecture series. ITAS is developed at the Center for Real-Time Computing (CRTC), Old Dominion University. Gemini was used to improve readability across the article; the authors take full responsibility for all content.

\newpage

\bibliographystyle{IEEEtran}
\bibliography{ref}

\end{document}